\documentclass[sigconf]{acmart}

\newcommand{\red}[1]{#1}
\newcommand{\blue}[1]{#1}
\usepackage{tabularx}
\usepackage{multirow}
\usepackage{multicol}
\usepackage{textcomp}
\usepackage{enumitem}
\usepackage{subfigure}
\usepackage{bm}
\usepackage[normalem]{ulem}
\useunder{\uline}{\ul}{}
\usepackage[ruled,vlined]{algorithm2e} 
\usepackage[switch]{lineno} 

\AtBeginDocument{%
  \providecommand\BibTeX{{%
    \normalfont B\kern-0.5em{\scshape i\kern-0.25em b}\kern-0.8em\TeX}}}

\copyrightyear{2023}
\acmYear{2023}
\setcopyright{acmcopyright}\acmConference[WSDM '23]{Proceedings of the
Sixteenth ACM International Conference on Web Search and Data
Mining}{February 27-March 3, 2023}{Singapore, Singapore}
\acmBooktitle{Proceedings of the Sixteenth ACM International Conference on
Web Search and Data Mining (WSDM '23), February 27-March 3, 2023,
Singapore, Singapore}
\acmPrice{15.00}
\acmDOI{10.1145/3539597.3570379}
\acmISBN{978-1-4503-9407-9/23/02}

\begin{document}

\title{One for All, All for One: Learning and Transferring User Embeddings for Cross-Domain Recommendation}
\author{Chenglin Li}
\affiliation{
  \institution{University of Alberta}
  \city{Edmonton, AB}
  \country{Canada}
}

\author{Yuanzhen Xie}
\affiliation{
  \institution{Platform and Content Group, Tencent}
  \city{Shenzhen}
  \country{China}
}

\author{Chenyun Yu}
\affiliation{
  \institution{Sun Yat-sen University}
  \city{Shenzhen}
  \country{China}
}

\author{Bo Hu}
\affiliation{%
  \institution{Platform and Content Group, Tencent}
  \city{Shenzhen}
  \country{China}
}

\author{Zang Li}
\affiliation{
  \institution{Platform and Content Group, Tencent}
  \city{Shenzhen}
  \country{China}
}

\author{Guoqiang Shu}
\affiliation{%
  \institution{Platform and Content Group, Tencent}
  \city{Shenzhen}
  \country{China}
}

\author{Xiaohu Qie}
\affiliation{%
  \institution{Platform and Content Group, Tencent}
  \city{Shenzhen}
  \country{China}
}

\author{Di Niu}
\affiliation{
  \institution{University of Alberta}
  \city{Edmonton, AB}
  \country{Canada}
}


\begin{abstract}
\label{sec:abstract}
Cross-domain recommendation is an important method to improve recommender system performance, especially when observations in target domains are sparse. However, most existing techniques focus on single-target or dual-target cross-domain recommendation (CDR) and are hard to be generalized to CDR with multiple target domains. In addition, the negative transfer problem is prevalent in CDR, where the recommendation performance in a target domain may not always be enhanced by knowledge learned from a source domain, especially when the source domain has sparse data. In this study, we propose CAT-ART, a multi-target CDR method that learns to improve recommendations in all participating domains through representation learning and embedding transfer. Our method consists of two parts: a self-supervised \textbf{C}ontrastive \textbf{A}u\textbf{T}oencoder (CAT) framework to generate global user embeddings based on information from all participating domains, and an \textbf{A}ttention-based \textbf{R}epresentation \textbf{T}ransfer (ART) framework which transfers domain-specific user embeddings from other domains to assist with target domain recommendation. CAT-ART boosts the recommendation performance in any target domain through the combined use of the learned global user representation and knowledge transferred from other domains, in addition to the original user embedding in the target domain. We conducted extensive experiments on a collected real-world CDR dataset spanning 5 domains and involving a million users. Experimental results demonstrate the superiority of the proposed method over a range of prior arts. We further conducted ablation studies to verify the effectiveness of the proposed components. Our collected dataset will be open-sourced to facilitate future research in the field of multi-domain recommender systems and user modeling.
\end{abstract}




\maketitle

\section{Introduction}
\label{sec:intro}
Data sparsity is a long-standing issue hampering traditional single-domain recommendation systems. 
Cross-domain Recommendation (CDR) has proven to be a promising approach to alleviate the data sparsity issue, aiming to leverage auxiliary information learned from other domains to improve recommendation in the target domain.
However, most prior research on cross-domain recommendation focuses on either the single-target CDR (STCDR) or dual-target CDR (DTCDR)~\cite{zhu2021cross} scenarios, with only two domains involved. STCDR aims to improve the recommendation accuracy in the target domain, while DTCDR tries to improve the performance in both domains simultaneously. 
Multi-target CDR (MTCDR) is a more general and challenging problem, which aims to improve the recommendation performance in multiple participating domains concurrently. 
Since previous methods solve the STCDR and DTCDR by modeling a pair-wise domain-domain relationship, 
intuitively speaking, extending them to the MTCDR scenario with $n$ domains involves handling at least ${n \choose 2}$ pairs of relations, which is not practical when the number of domains is large.

Relatively fewer efforts have been put onto MTCDR.
Current state-of-the-art solutions usually generate a shared cross-domain user representation for each user, which, combined with domain-specific features, is used to boost recommendations in any given domains~\cite{cui2020herograph,yan2021multi}.  
HeroGRAPH~\cite{cui2020herograph} collects user behavior in all domains to build a heterogeneous graph. It then applies the graph convolutional networks (GCN)~\cite{hamilton2017inductive} to generate the cross-domain user and item embeddings, which are directly transferred to target domains to boost recommendation. 
However, most recommender systems are built on users' sensitive data, e.g., check-in data, browsing records, which is held by different domains and cannot be shared directly to form a large heterogeneous graph. 
MPF~\cite{yan2021multi} learns a global embedding for each user, which is directly shared among all domains and optimized by the recommendation losses in all participating domains via multi-task learning. 
However, the cross-domain user representation, extracted directly on the collected data from all domains, may be severely biased by the domains with richer data and may fail to model the user preferences in sparse domains. 
The biased global representation of a user may negatively affect recommendation performance when transferred to a target domain. This unbalanced data problem also exists in HeroGRAPH~\cite{cui2020herograph}.

We observe that two types of user embeddings can both be helpful in cross-domain recommendation, including 1) the global user embedding, which represents the overall domain-invariant characteristics of a user, and 2) domain-specific user embeddings, which model the user behavior in various individual domains. 
Two questions naturally arise concerning the interaction of these embeddings. First, can we generate a global user representation only based on the user's embeddings obtained from individual domains? 
If this global user embedding is representative enough and not biased toward a single domain, it can be directly used to improve recommendation in all domains. 
We refer to the first goal by ``One for All", as in One global user embedding for recommendations in All domains. To ensure better data isolation, ideally, the global user embedding should be synthesized only based on domain-specific user embeddings without accessing raw behavior data in each domain.

On the other hand, what is missing in prior work on MTCDR \cite{cui2020herograph, yan2021multi} is the transfer of domain-specific user embeddings to assist with recommendation in the target domain. However, directly transferring these features may cause performance degradation in the target domain due to various reasons, e.g., low-quality embeddings transferred from irrelevant domains. This phenomenon is often referred to as negative transfer \cite{zhang2020overcoming}.
The second question we ask is--can we also transfer domain-specific user embeddings in one domain to help improve prediction in another domain while avoiding negative transfer? 
We refer to the second goal by ``All for One'', as in All domain-specific user embeddings for helping the recommendation in One domain. 

To address the aforementioned challenges, we propose CAT-ART, a novel multi-target CDR framework, which starts from using traditional matrix factorization to pretrain domain-specific user embeddings in each domain.
To achieve the concept of One for All, we propose a Contrastive AuToencoder (CAT) to learn a global user embedding solely based on the pretrained domain-specific user embeddings from all the domains, without directly accessing raw behavior data in each domain. 
To attain the goal of All for One, we build an Attention-based Representation Transfer (ART) unit in each target domain, which transfers and utilizes the pretrained domain-specific embeddings to boost its recommendation performance while minimizing the impact of negative transfer.
Our contributions can be summarized as follows:

We introduce a contrastive autoencoder (CAT), which learns a general global embedding for each user by reconstructing the concatenated sequence of domain-specific user embeddings. To further benefit from self-supervised representation learning \red{and extract robust representations from domain-specific embeddings}, we randomly mask some domain in the input sequence to the autoencoder and learn to reconstruct the original sequence, while a contrastive self-supervised loss is used to ensure the masked and unmasked domain-specific user embedding sequences can map to similar global user embeddings for the same user. 

We propose Attention-based Representation Transfer (ART), which judiciously adapts the domain-specific user embeddings from other domains to the target domain according to an attention mechanism. ART then combines the target domain user embedding, the global user embedding, as well as the adapted domain-specific user embeddings to jointly improve recommendation in the target domain.
    
\red{To evaluate our method in real scenarios, we have collected a large dataset involving over a million users spanning 5 domains, including App installation (App-install), Recent App usage (App-use), article viewing, short-video viewing, and long-video viewing. Each of these domains has its own user behavior history and independently pretrained user embeddings. 
We conduct extensive experiments on the collected data and demonstrate the superiority of the proposed CAT-ART method by comparing with a range of state-of-the-art MTCDR baselines.
Experimental results suggest that CAT-ART significantly outperforms all baselines in most of the domains, e.g., App-install, App-use, article, long-video, on several evaluation metrics. Moreover, we show that while other state-of-the-art baselines are severely impacted by negative transfer, CAT-ART can effectively avoid the negative transfer issue. 
}


\section{Related Work}
\label{sec:related}

\textbf{Single-Target CDR}. 
Previous single-target CDR (STCDR) works can be classified into content-based \cite{elkahky2015multi,kanagawa2019cross,yuan2019darec} and embedding-based approaches\cite{man2017cross, kang2019semi, wang2019solving,zhao2020catn}.
In the former,
\citeauthor{kanagawa2019cross}~\cite{kanagawa2019cross} propose a content-based domain adaptation model and a domain separation network for cross-domain recommendations. 
In the latter, EMCDR \cite{man2017cross} solves the cold-start problem in the target domain by learning a mapping function between the user embeddings of the source and the target domain. 
CDIE-C \cite{wang2019solving} enhances item embedding learning by cross-domain co-clustering for the sequential recommendation.

\textbf{Dual-Target CDR}. 
Given two domains, DTCDR is to improve the recommendation accuracy in both domains at the same time by leveraging their observed information\cite{zhu2019dtcdr,liu2020cross,zhu2020graphical,li2020ddtcdr,li2021dual,10.1145/3488560.3498388,zhu2022personalized}. 
\citeauthor{zhu2019dtcdr}~\cite{zhu2019dtcdr} first proposed the DTCDR problem and a DTCDR framework that learns more representative embeddings of users and items based on multi-sources. 
\citeauthor{liu2020cross}~\cite{liu2020cross} combine the embeddings of common users based on hyper-parameters and data sparsity degrees of users. 
GA-DTCDR \citeauthor{zhu2020graphical} employs graph embedding to generate more informative embeddings of users and items, and employs element-wise attention to combine the embeddings of common users/items across domains. 
The deep dual transfer CDR (DDTCDR) \cite{li2020ddtcdr} considers the bi-directional latent relations between users and items and applies a latent orthogonal mapping to extract user preferences. 
CATN~\cite{zhao2020catn} learns aspect correlations across domains with an attention mechanism. 
Some work focuses on extracting domain-invariant or domain-independent user attributes for CDR~\cite{zhuang2018cross,liu2020exploiting,sahu2020knowledge}. ACDN~\cite{liu2020exploiting} models individual’s propensity from the aesthetic perspective and captures users’ domain-independent aesthetic preference for CDR.

\textbf{Multi-Target CDR}.
The MTCDR methods \cite{cui2020herograph, krishnan2020transfer, yan2021multi, kim2019domain, yuan2020parameter,yuan2021one} have emerged to improve the recommendation performance of multiple domains simultaneously.
\blue{\citeauthor{kim2019domain} \cite{kim2019domain} adopt recurrent neural networks (RNNs) to model the sequential behavior of users in multiple domains simultaneously. \cite{yuan2020parameter,yuan2021one} achieve knowledge transfer through parameter sharing across multiple domains. 
However, they all focus on sequential recommendation.}
Recent works try to extract domain-specific and cross-domain features simultaneously, e.g., MSDCR~\cite{zhao2022multi}, HeroGRAPH~\cite{cui2020herograph}, and MPF~\cite{yan2021multi}. 
Specifically,   
HeroGRAPH~\cite{cui2020herograph} constructs heterogeneous graph from interactions between users and items from all domains and develops a graph embedding algorithm to extract common features for MTCDR.
MPF~\cite{yan2021multi} captures both the cross-site and site-specific preferences for multi-site video recommendations. 
\red{GA-MTCDR~\cite{zhu2021unified}, extended from GA-DTCDR~\cite{zhu2020graphical}, employs element-wise attention to combine embeddings of overlapped users/items from all domains. However, it requires side information for graph construction in each domain.} 
However, 
\blue{most of the previous MTCDR methods ignore the data isolation constraint between domains in practice, and none of them has considered the negative transfer problem. In this study, we try to solve the MTCDR problem in a more realistic scenario where user and item data are held by each individual domain and cannot be shared across domains. Furthermore, our framework is also designed to avoid the negative transfer problem in MTCDR.}



\section{Methodology}
\label{sec:sys}

\begin{figure*}[ht]
  \centering
  \includegraphics[width=5.2in]{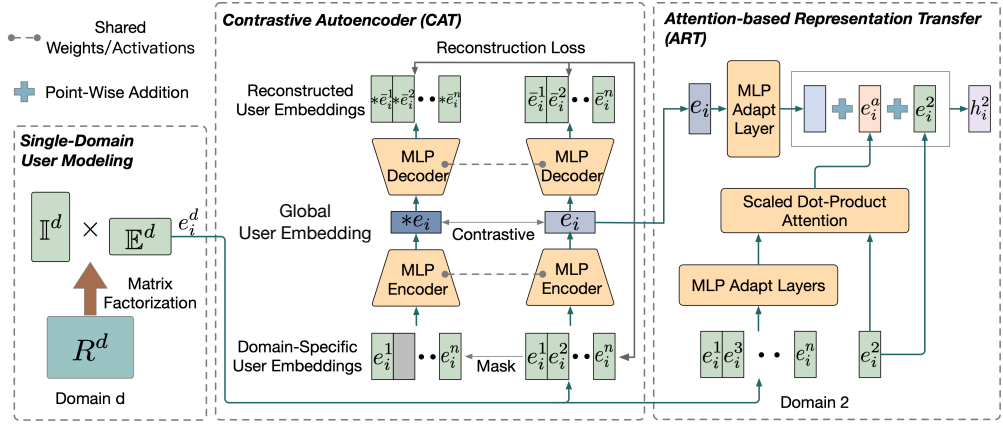}
  \caption{
  The architecture of the CAT-ART model. The CAT module takes domain-specific user embeddings as input and generates global user representation in a self-supervised manner. Then, the global user embedding $e_i$ and the domain-specific embeddings from all the other domains are transferred to a target domain, e.g., domain 2, for boosted recommendations.
  }
  \Description{A figure shows the overall structure of the CAT-ART model.}
  \label{fig:sys}
\end{figure*}


We focus on the MTCDR problem with a global user set $U$, and item sets $\{V_1, \cdots V_n\}$ in $n \ge 3$ domains. 
User-item interactions in domain $d \in [1, \cdots, n]$ are represented by a matrix $R^d$ with the shape of $|U| \times |V_d|$ where $|U|$ is the number of users and $|V_d|$ denotes the number of items in domain $d$. 
\blue{Under} the implicit feedback setting, \blue{all the elements in the matrix $R^d$ are with the value of either 1 or 0}, which indicates whether there is an interaction between a given user-item pair, i.e., for user $i$ and item $j$ in domain $d$, $r^{d}_{ij} \in [0, 1] , \forall{i} \in |U|, \forall{j} \in |V_d|$. 
We further consider the \blue{scenario} of data isolation in practical applications, that is, the interactive information between users and items in a specific domain is not observable by other domains.
Our goal is to improve the recommendation accuracy in all $n$ domains simultaneously based on the interaction matrices. 

\subsection{Architecture Overview of CAT-ART}
We set two objectives when dealing with the multi-target CDR problem.
That is, 1) One for All: extracting global user representation that is used for recommendations in all domains. 
And 2) All for One: transferring domain-specific embeddings from all available domains to assist the recommendation in a target domain without negative transfer. 
To achieve these two objectives and avoid the direct use of raw data across domains, we propose the CAT-ART model where the Contrastive Autoencoder (CAT) module and the Attention-based Representation Transfer (ART) unit are designed for the above two objectives, respectively. 

Figure~\ref{fig:sys} provides an overview of the CAT-ART model.
\red{First, the domain-specific user embeddings are pretrained within each domain independently using BPRMF~\cite{rendle2012bpr}.}
Then, the CAT module takes the domain-specific embeddings collected from all domains as input and generates global user representation. 
To create unbiased cross-domain user embeddings, we combine the reconstruction loss and contrastive self-supervised loss in model training, so that the CAT module is capable of extracting informative global user representation.
Finally, the ART module transfers the domain-specific user embeddings from all the other domains to boost the recommendation performance in a single domain, e.g., domain 2 in Figure~\ref{fig:sys}. 
\red{By incorporating attention mechanisms to the ART module, the contribution of each domain can be judiciously adjusted according to their relevance so as to address the negative transfer issue.}

\subsection{Domain-specific User Embedding}
\red{Shown in Figure~\ref{fig:sys}, in the Single-Domain User Modeling unit, }
we adopt the widely used Matrix Factorization (MF) model~\cite{ma2008sorec} with the Bayes Personalized Ranking (BPR)~\cite{rendle2012bpr} loss to get the user and item embeddings in each domain. 
Formally, to factorize the interaction matrix $R^d$ in domain $d$, we create two trainable embedding matrices $\mathbb{I}^d \subset \mathbb{R} ^{(|V_d| + 1)\times m}$ and $\mathbb{E}^d \subset \mathbb{R}^{(|U| + 1) \times m}$ to represent item and user embeddings in domain $d$, respectively. Where $m$ represents the number of dimensions in the latent space. 
For simplicity, we set the embedding dimensions of users and items in all domains to be $m$.
In domain $d$, given an user $u_i$ with embedding $\bm{e}^d_{i} \in \mathbb{E}^d$ and an item $v^d_j \in V^d$ with embedding $\bm{I}^d_{j} \in \mathbb{I}^d$, the preference score of the user to the item is computed as the inner product between their embeddings , $r^d_{ij} = \bm{e}^d_{i} \bm{I}^d_{j}$. Note that, we use bold font to denote vector variables.
Then, the BPR loss in domain $d$ is formulated as:
\begin{equation} \label{eq:loss_rc}
\mathcal{L}_{bpr}^d = - \sum_{i \in U} \sum_{j \in p^d_{i}} \sum_{l \notin p^d_{i}} \log\sigma(r^d_{ij} - r^d_{il}),
\end{equation}
where $p^d_i$ is the set of items that user $i$ has interacted with in domain $d$, and $\sigma(\cdot)$ represents the sigmoid function.  

\red{By minimizing the BPR loss, we obtain domain-specific user embeddings suitable for recommendation in each domain. 
Instead of sharing raw user data, we only share the pretrained domain-specific user embeddings across domains to enable knowledge sharing across domains under the raw data isolation constraint.}

\subsection{Contrastive Autoencoder}
The pretrained domain-specific user embeddings are collected and fed into the CAT module to get global user representations. 
To do this, we adopt an autoencoder framework which takes the domain-specific embeddings of a user as input, and generates latent user representation.   
Specifically, we first stitch all domain-specific user embeddings, which is a set of real-valued dense vectors, into a large one-dimensional vector in a predefined order, e.g., from domain 1 to domain n. Then, an encoder is used to extract the latent user presentation which is further fed into a decoder to reconstruct the input domain-specific embeddings. 
We use Multi-Layer Perception (MLP)~\cite{gardner1998artificial} to build both the encoder and decoder modules.
Formally, for a given user $u_i$ with its pretrained domain-specific embeddings $\{\bm{e}^1_i, \bm{e}^2_i, \cdots, \bm{e}^{n}_i\}$, we apply the following procedures to get its latent representation: 
\begin{equation} \label{eq:cross-domain}
\begin{aligned} 
\bm{e}_i &= \textrm{ML}P_{\textrm{enc}}(\bm{e}_i^{1\frown} \bm{e}_i^{2\frown} \cdots ^\frown \bm{e}_i^n)            \\   
\bar{\bm{e}}_i^{1\frown} \bar{\bm{e}}_i^{2\frown} \cdots ^\frown \bar{\bm{e}}_i^n &= \textrm{MLP}_{\textrm{dec}}(\bm{e}_i), \\
\end{aligned}
\end{equation} 
where $\bm{e}_i$ is the latent representation of user $u_i$, and 
$\bm{e}_i^{1\frown} \bm{e}_i^{2}$ denotes the long one-dimensional vector after concatenating the vectors $\bm{e}_i^{1}$ and $ \bm{e}_i^{2}$. 
Here $\bar{\bm{e}}_i^{d}$ represents the reconstruction of user embedding $\bm{e}_i^{d}$ in domain $d$. Note that the latent vector $\bm{e}_i$ has the same size as domain-specific embeddings in individual domains, i.e., $|\bm{e}_i| = |\bm{e}_i^d| = m, \forall{d} \in [1, n] $.

By optimizing the mean square reconstruction error~\eqref{eq:rec_loss} between the input and the reconstructed embeddings, the encoder can learn the most important global attributes to reconstruct the domain-specific user embeddings in each domain. 
\begin{equation} \label{eq:rec_loss}
\begin{aligned} 
\mathcal{L}_{rec} = \frac{1}{|U|}\sum_{i \in U} \sum^n_{d = 1} ||\bm{e}^d_i - \bar{\bm{e}}^{d}_i ||_2. 
\end{aligned}
\end{equation} 

\red{On the other hand, the effectiveness of the pretrained domain-specific user embeddings, which are the input of the autoencoder, is highly affected by the data quality and sparsity in each domain. For example, under-trained user embeddings from a sparse domain may introduce noise to the input of the autoencoder. Furthermore, the autoencoder may be biased towards domains with a higher quality of user embedding, as it is easier to reconstruct a well-trained embedding than an under-trained embedding with noise.} 

Therefore, we adopt the contrastive self-supervised learning~\cite{chen2020simple} to further train the autoencoder for a more general and robust latent user representation that does not bias to any specific domain.
\red{The core idea of contrastive self-supervised learning is to make the representation of an input sample agree with that of an augmented sample, e.g., obtained by applying Gaussian noise or  Cutout~\cite{chen2020simple}.} 

In our problem, contrastive self-supervised learning is integrated into the autoencoder framework to extract the global representations of users from their domain-specific embeddings. 
We adopt an \emph{Mask} operation to generate the ``augmentations'' for an input, e.g., $\bm{e}_i^{1\frown} \bm{e}_i^{2\frown} \cdots ^\frown \bm{e}_i^n$. 
Specifically, with \emph{Mask}, we generate ``augmented'' inputs by removing the domain-specific user embeddings from several randomly selected domains. 
Formally, the following procedure is applied:
\begin{equation} \label{eq:contrastive}
\begin{aligned} 
\bm{e}_i^{1\frown} \bm{e}_m^{\frown} \cdots ^\frown \bm{e}_i^n &= \textrm{Mask}(\bm{e}_i^{1\frown} \bm{e}_i^{2\frown} \cdots ^\frown \bm{e}_i^n) \\
\ast \bm{e}_i &= \textrm{MLP}_{\textrm{enc}}(\bm{e}_i^{1\frown} \bm{e}_m^{\frown} \cdots ^\frown \bm{e}_i^n)
\end{aligned}
\end{equation} 
where $\bm{e}_m$ is a trainable vector used to replace the domain-specific embeddings of masked domains, e.g., domain 2 in Figure~\ref{fig:sys}. 
After random masking and padding, the ``augmented'' sample, i.e., $\bm{e}_i^{1\frown} \bm{e}_m^{\frown} \cdots ^\frown \bm{e}_i^n$, is fed into the encoder \emph{$\textrm{MLP}_{\textrm{enc}}$} to get the latent representation $\ast \bm{e}_i$. 
Then, a contrastive loss function is defined for the contrastive prediction task. Given a set of latent embeddings $\{\bm{e}_1, \cdots \bm{e}_k, \ast \bm{e}_1, \cdots, \ast \bm{e}_k\}$ where $(\bm{e}_i, \ast \bm{e}_i)$ forms a positive pair of representations, the contrastive prediction is to identify $\ast \bm{e}_i $ in $\{\ast \bm{e}_1, \cdots, \ast \bm{e}_k \}$ for a given $ \bm{e}_i $, and vice versa, to identify $\bm{e}_i $ in $\{\bm{e}_1, \cdots, \bm{e}_k \}$ for a given $ \ast  \bm{e}_i $. 

We randomly sample a minibatch of $N$ users and define the contrastive autoencoder task on the pairs of ``augmented'' and original user embeddings derived from the minibatch, resulting in $2N$ representations. Let $\phi(\bm{e}_i, \ast \bm{e}_i) = \frac{\bm{e}_i \ast \bm{e}_i^T}{|\bm{e}_i|| \ast \bm{e}_i|} $
denotes the cosine similarity between $\bm{e}_i$ and $\ast\bm{e}_i$. Then, the contrastive loss of user $i$ in the minibatch is computed as:
\begin{equation} \label{eq:contrastive_loss}
\begin{aligned} 
l_{i} = -log \frac{exp(\phi(\bm{e}_i, \ast \bm{e}_i)/\tau)}{\sum_{k=1}^N exp({\phi(\bm{e}_i, \ast \bm{e}_k})/\tau)} - log \frac{exp(\phi(\ast \bm{e}_i, \bm{e}_i)/\tau)}{\sum_{k=1}^N exp({\phi(\ast \bm{e}_i, \bm{e}_k})/\tau)}, 
\end{aligned}
\end{equation} 
where $\tau$ is a temperature parameter. The first term defines the contrastive prediction loss when identifying $\ast e_i$ given $e_i$ and the second term is the loss for identifying $e_i$ given $\ast e_i$.  
\red{By minimizing the contrastive loss, the encoder is trained to put the embeddings of a user and its ``augmentations'', i.e., $e_i$ and $\ast e_i$, close to each other in the latent embedding space.
By doing this, we can extract a more general latent user representation that is robust to noisy input introduced by low-quality domain-specific user embeddings from a sparse domain.
}

Apart from the contrastive loss, we further put the $\ast e_i$ into the decoder $\textrm{MLP}_{\textrm{dec}}$ to reconstruct the original input.  
That is, instead of minimizing the reconstruction error between the ``augmentation'', i.e., $\bm{e}_i^{1\frown} \bm{e}_m^{\frown} \cdots ^\frown \bm{e}_i^n$, and its reconstruction, i.e., $\ast \bar{\bm{e}}_i^{1\frown} \ast \bar{\bm{e}}_i^{2\frown} \cdots ^\frown \ast \bar{\bm{e}}_i^n $, we optimize the reconstruction loss between the original domain-specific embeddings and the reconstructed embeddings from $\textrm{MLP}_{\textrm{dec}}(\ast \bm{e}_i)$, that is,  
\begin{equation} \label{eq:rec_loss2}
\begin{aligned} 
\ast \bar{\bm{e}}_i^{1\frown} \ast \bar{\bm{e}}_i^{2\frown} \cdots ^\frown \ast \bar{\bm{e}}_i^n &= \textrm{MLP}_{\textrm{dec}}(\ast \bm{e}_i), \\
\mathcal{L}_{rec}^\ast &= \frac{1}{|U|}\sum_{i \in U} \sum^n_{d = 1} ||e^d_i - \ast \bar{\bm{e}}^{d}_i ||_2.
\end{aligned}
\end{equation} 
By minimizing $\mathcal{L}_{rec}^\ast$, the encoder can extract latent representation which is informative enough to reconstruct the masked domains even when the embeddings of masked domains are missing. 
\red{This further encourages the encoder to extract unbiased global representations of users. }
Finally, the loss function to train the CAT module can be summarized as:
\begin{equation} \label{eq:total_loss}
\begin{aligned}  
\mathcal{L}_{cat} = \alpha_1 \mathcal{L}_{rec} + \alpha_2 \mathcal{L}_{rec}^\ast + (1- \alpha_1 - \alpha_2) \sum_{i=1}^{|U|}l_i, 
\end{aligned}
\end{equation} 
where $\alpha_1$ and $\alpha_2$ are hyper-parameters to control the weights of each component. The third term represents the sum of the contrastive loss over all users.

The extracted global representation represents the general preferences and overall characteristics of users, that are beneficial to the recommendation task in each domain. 
In addition, because the global user representation is not biased to any domain, it can be directly transferred to any target domain without worrying about the negative transfer issue.

\subsection{Attention-based Representation Transfer} 
The domain-specific embeddings represent the users' preferences in different domains, which are useful features to boost recommendations in a specific domain. 
However, the direct transfer of domain-specific user embeddings is prone to negative transfers in the target domain. 
\red{First, if two domains are unrelated, the transferred embeddings may have little or even negative impact on the target domain. Second, if a domain-specific user embedding is under-trained in its original domain due to data sparsity or insufficient data quality, directly transferring such an embedding may introduce noise to the target domain, leading to performance degradation.} 

Therefore, we build an attention-based representation transfer (ART) unit in each domain, to integrate the domain-specific embeddings from other domains for better recommendations in the current domain. 
Specifically, 
we first build an MLP-based domain adaptation layer for domain-specific embeddings from each domain. 
Then, we adopt the scaled-dot produce attention \cite{vaswani2017attention} which uses the user embedding in the target domain as query and attends to the domain-specific embeddings from other domains.  
Formally, in domain $d$, we have  
\begin{equation} \label{eq:attention}
\begin{aligned}  
Q &= \bm{e}^d_i \\
K = V &= \textrm{MLP}_{\textrm{adapt}}(\{\bm{e}^k_i\}, k\ne d )\\
\bm{e}^{a}_i = Attention(Q, K, V) &= \textrm{softmax}(\frac{Q K^T}{\sqrt{m}})V,
\end{aligned}
\end{equation} 
where $m$ is the dimensionality of the user embedding $\bm{e}^d_i$ in domain $d$. 
\red{With attention, the ART module can assign more weights to the embeddings from the most related domains and reduce the influence of noisy embeddings from sparse or unrelated domains, and thus can effectively alleviate negative transfers in MTCDR.}

\textbf{Final User Embedding}. 
We build a MLP-based domain adaptation module, i.e., \emph{$\textrm{MLP}_{\textrm{ind}}(\cdot)$}, which adapts the global embedding to current domain for recommendation. 
Then, we combine features from \blue{the global and domain-specific user embeddings} with point-wise addition. For example, in domain $d$, we have 
\begin{equation} \label{eq:final_emb}
\begin{aligned}  
\bm{h}^d_i = \bm{e}^d_i + \textrm{MLP}_{\textrm{ind}}(\bm{e}_i) + \bm{e}^a_i, 
\end{aligned}
\end{equation} 
where the first term is the user embedding in current domain, the second term represents the information transferred from the global user representation, and the last term denotes features from all the other domains.  
With user embedding $h^d_i$, the preference score of the user $i$ to an item $j$ in domain $d$ is recalculated as: $r^d_{ij} = h^d_i I_j^d$. 

\subsection{\red{Model Training}}
We adopt the following three steps to train the CAT-ART model. 
First, we apply the BPRMF model in individual domains to get the pretrained domain-specific user and item embeddings based on BPR loss~\eqref{eq:loss_rc}. 
Then, 
we train the CAT module in a self-supervised manner with both the reconstruction loss and contrastive loss~\eqref{eq:total_loss}. 
Finally, we fix the domain-specific user embeddings and the CAT module, and optimize the parameters of the \emph{$\textrm{MLP}_{\textrm{ind}}(\cdot)$} and ART module according to the BPR loss \blue{in individual domains.} 


\section{Experiments}
\label{sec:exp}
We conduct extensive experiments on a collected dataset and compare our approach with state-of-the-art MTCDR methods. 

\subsection{Datasets}
\begin{table}[tbp]
\caption{Statistics of the Collected Dataset with 5 Domains.}  
\begin{center}
\begin{tabular}{lcccc}
\toprule
Domain  & \#Users                     & \#Items     & \#Interactions & Density(\textperthousand)    \\
\hline
App-Ins & \multirow{5}{*}{1,166,552}    & 100,000     & 101,981,793   & 0.874     \\
APP-Use &                             & 100,000     & 18,156,535    & 0.155     \\
Articles &                            & 50,000      & 102,832,656   & 1.763     \\
Video-S &                             & 50,000      & 74,911,020    & 1.284     \\
Video-L &                             & 50,000      & 11,412,988    & 0.196     \\    
\bottomrule
\end{tabular}
\label{table:dataset_info}
\end{center}
\end{table}

We collected user logs from five domains through multiple real apps \red{owned by Tencent}. 
Specifically, they are Application installation preferences (``APP-Ins''), Application usage preferences (``APP-Use''), ``Articles'', Short Video (``Video-S''), and Long Video (``Video-L''). 
Among them, the ``APP-Ins'' and ``APP-Use'' data is collected from January 2021 to August 20, 2021, and the data of ``Articles'', ``Video-S'', and ``Video-L'' are collected from June 2021 to August 2021. 
\blue{
We use the tags that best describe the visited items to avoid a tremendous set of items.}
TF-IDF~\cite{ramos2003using} is applied to select a fixed number of the most informative tags in each domain, resulting in 100,000 tags in the ``APP-Ins'' and ``APP-Use'' domains and 50,000 tags in the rest of domains.
\red{We filter out users with less than 5 visited tags.  
Furthermore, the number of tags a user has visited in each domain is truncated to a fixed number, i.e., a user can have a maximum of 100 tags in the ``Video-S'' domain, and a maximum of 300 tags in the other domains.}
\blue{
All the above data processing procedures
are provided by the data owner (Tencent), and have proven to be effective in real-world tasks such as gender and age prediction.} 
Table~\ref{table:dataset_info} provides detailed statistics of the collected dataset with 5 domains.

\subsection{Experimental Setup}
    
Our experiments aim at answering the following questions: 
\begin{itemize} [leftmargin=*]
    \item {\textbf{RQ1.}} How does CAT-ART perform vs. state-of-the-art baselines in the MTCDR task? 
    \item {\textbf{RQ2.}} Does our model handle the problem of negative transfer? 
    \item {\textbf{RQ3.}} How does the sub-modules help the model succeed in solving the MTCDR problem? 
\end{itemize}
To train and evaluate a model, we randomly split the interactions of a user in each domain into three parts for: training (70\%), validation (10\%), and testing (20\%). 
\blue{The evaluation metrics are Precision@K, Recall@K, and NDCG@K~\cite{jarvelin2002cumulated}
that are computed by the all-ranking protocol where all items/tags are ranked. 
}
We repeat all experiments three times and give the average and standard deviation of all metrics.

\noindent \textbf{Compared Methods}. 
We compare our model with single-domain recommendation methods(SMF), STCDR methods(CMF), and MTCDR methods(MPF, GA-MTCDR, HeroGRAPH).
\begin{itemize} [leftmargin=*]
    \item \textbf{SMF}: It factorizes the user-tag interaction matrix of each domain separately based on the BPR loss~\cite{rendle2012bpr}.
    
    \item \textbf{CMF}~\cite{singh2008relational}: It collects interactions from all domains to form a single matrix which is further factorized for recommendation.
    
    \item \textbf{HeroGRAPH-L}: HeroGRAPH~\cite{cui2020herograph} learns the cross-domain and domain-specific representations. 
    Since the HeroGRAPH does not open-source its code, we implement this method based on Lightgcn~\cite{he2020lightgcn}.
    
    \item \textbf{MPF}~\cite{yan2021multi}: It captures the cross-domain preference with user's behavior in all domain, \red{and combine it with the user embedding in the target domain for recommendation. }
    
    \item \red{\textbf{GA-MTCDR}~\cite{zhu2021unified}: It adopts the node2vec~\cite{grover2016node2vec} model to pretrain the user/item embeddings in each domain, and uses element-wise attention to transfer embeddings among multiple domains.
    }

\end{itemize}

\begin{table*}[!ht]
\begin{center}
\caption{
Results (in \%) of the Proposed Method and Baselines. The $\downarrow$ represents negative transfer compared with SMF.}
\resizebox{0.75\textwidth}{!}{
\begin{tabular}{c|l|cc|cc|cc}

\toprule
\multirow{2}{*}{Model}     & \multirow{2}{*}{Domain}  & \multicolumn{2}{c|}{Precision} & \multicolumn{2}{c|}{Recall}  & \multicolumn{2}{c}{NDCG}    \\
\cline{3-8}                          
                      &    & @10   & @20   & @10   & @20   & @10   & @20   \\
\midrule
\multirow{5}{*}{SMF}       & APP-Ins & 33.82$\pm{0.70}$  & 25.46$\pm{0.88}$ & 21.51$\pm{0.39}$ & 31.91$\pm{1.22}$   & 32.56$\pm{0.43}$ & 32.53$\pm{0.89}$ \\
                           & APP-Use & 20.91$\pm{0.23}$  & 12.21$\pm{0.26}$ & 65.5$\pm{0.89}$  & 75$\pm{1.50}$      & 57.39$\pm{1.46}$ & 60.81$\pm{1.72}$ \\
                           & Article & 16.02$\pm{0.73}$  & 12.05$\pm{0.58}$ & 16.64$\pm{0.43}$ & 23.25$\pm{0.40}$   & 21.59$\pm{1.30}$ & 21.93$\pm{1.06}$ \\
                           & Video-S & 3.9$\pm{0.03}$    & 3.86$\pm{0.02}$  & 3.59$\pm{0.44}$  & 6.9$\pm{0.77}$     & 3.83$\pm{0.13}$  & 4.84$\pm{0.25}$  \\
                           & Video-L & 5.98$\pm{0.20}$   & 3.91$\pm{0.10}$  & 26.73$\pm{0.87}$ & 34.6$\pm{0.88}$    & 20.37$\pm{1.19}$ & 22.91$\pm{1.2}$  \\
\hline                           
\multirow{5}{*}{CMF}       & APP-Ins & 33.57$\pm{0.37}^{\downarrow}$ & 25.19$\pm{0.37}^{\downarrow}$ & 21.8$\pm{0.19}$ & 32.05$\pm{0.43}$ & 32.39$\pm{0.29}^{\downarrow}$ & 32.45$\pm{0.27}^{\downarrow}$  \\
                           & APP-Use & 20.41$\pm{0.11}^{\downarrow}$ & 12.17$\pm{0.05}^{\downarrow}$ & 64.91$\pm{0.27}^{\downarrow}$ & 75.54$\pm{0.16}$ & 43.99$\pm{0.78}^{\downarrow}$ & 47.89$\pm{0.74}^{\downarrow}$  \\
                           & Article & 10.29$\pm{0.27}^{\downarrow}$ & 8.37$\pm{0.19}^{\downarrow}$  & 8.83$\pm{0.23}^{\downarrow}$  & 13.79$\pm{0.28}^{\downarrow}$  & 11.24$\pm{0.34}^{\downarrow}$ & 12.07$\pm{0.31}^{\downarrow}$  \\
                           & Video-S & 3.87$\pm{0.12}$ & 3.81$\pm{0.12}^{\downarrow}$ & 4.08$\pm{0.17}$ & 7.6$\pm{0.29}$  & 4.00$\pm{0.14}$ & 5.04$\pm{0.18}$ \\
                           & Video-L & 4.74$\pm{0.03}^{\downarrow}$ & 3.26$\pm{0.01}^{\downarrow}$ & 21.44$\pm{0.12}^{\downarrow}$ & 29.14$\pm{0.06}^{\downarrow}$  & 12.67$\pm{0.07}^{\downarrow}$ & 15.14$\pm{0.05}^{\downarrow}$  \\
\hline                         
\multirow{5}{*}{MPF}       & APP-Ins & 36.08$\pm{1.53}$ & 27.11$\pm{0.41}$ & 23.28$\pm{0.99}$ & 34.29$\pm{0.44}$ & 36.95$\pm{5.56}$ & 36.53$\pm{4.02}$ \\
                           & APP-Use & 20.95$\pm{0.12}$ & 12.26$\pm{0.16}$ & 65.55$\pm{0.44}$ & 75.18$\pm{0.84}$ & 55.67$\pm{2.71}^{\downarrow}$ & 59.14$\pm{2.52}^{\downarrow}$  \\
                           & Article & 14.55$\pm{0.16}^{\downarrow}$ & 11.14$\pm{0.11}^{\downarrow}$ & 15.35$\pm{0.07}^{\downarrow}$ & 21.72$\pm{0.12}^{\downarrow}$ & 20.96$\pm{0.63}^{\downarrow}$ & 21.29$\pm{0.52}^{\downarrow}$  \\
                           & Video-S & 3.63$\pm{0.29}^{\downarrow}$  & 3.67$\pm{0.13}^{\downarrow}$  & 3.71$\pm{0.30}$ & 7.16$\pm{0.68}$ & 3.85$\pm{0.40}$ & 4.91$\pm{0.11}$ \\
                           & Video-L & 2.74$\pm{0.95}^{\downarrow}$  & 2.09$\pm{0.52}^{\downarrow}$  & 11.96$\pm{4.31}^{\downarrow}$ & 18.2$\pm{4.66}^{\downarrow}$ & 8.03$\pm{3.65}^{\downarrow}$ & 10.01$\pm{3.79}^{\downarrow}$ \\
\hline                           
\multirow{5}{*}{\red{GA-MTCDR}}  & APP-Ins & 16.77$\pm{0.05}^{\downarrow}$ & 10.35$\pm{0.02}^{\downarrow}$ & 11.7$\pm{0.01}^{\downarrow}$ & 14.37$\pm{0.03}^{\downarrow}$ & 17.81$\pm{0.08}^{\downarrow}$ & 16.01$\pm{ 0.03 }^{\downarrow}$  \\
                           & APP-Use & 13.88$\pm{0.05}^{\downarrow}$ & 10.46$\pm{0.01}^{\downarrow}$ & 45.44$\pm{0.13}^{\downarrow}$ & 67.2$\pm{0.16}^{\downarrow}$  & 32.35$\pm{0.13}^{\downarrow}$ & 40.16$\pm{0.1}^{\downarrow}$  \\
                           & Article & 4.62$\pm{0.13}^{\downarrow}$  & 3.73$\pm{0.03}^{\downarrow}$  & 4.12$\pm{0.14}^{\downarrow}$  & 6.37$\pm{0.11}^{\downarrow}$  & 6.22$\pm{0.18}^{\downarrow}$  & 6.36$\pm{0.13}^{\downarrow}$ \\
                           & Video-S & 3.44$\pm{0.03}^{\downarrow}$  & 3.1$\pm{0.02}^{\downarrow}$   & 3.48$\pm{0.08}^{\downarrow}$  & 6.03$\pm{0.06}^{\downarrow}$  & 4.22$\pm{0.05}$  & 4.69$\pm{0.04}$ \\
                           & Video-L & 3.18$\pm{0.15}^{\downarrow}$  & 2.22$\pm{0.07}^{\downarrow}$  & 14.21$\pm{0.74}^{\downarrow}$ & 19.76$\pm{0.54}^{\downarrow}$ & 10.46$\pm{0.63}^{\downarrow}$ & 12.23$\pm{0.49}^{\downarrow}$ \\
\hline                           
\multirow{5}{*}{HeroGRAPH-L}    & APP-Ins & 34.05$\pm{2.01}$ & 24.47$\pm{1.16}^{\downarrow}$ & 22.34$\pm{1.14}$ & 31.61$\pm{1.35}^{\downarrow}$ & 40.5$\pm{1.91}$ & 38.12$\pm{1.51}$  \\
                           & APP-Use & 20.68$\pm{0.36}^{\downarrow}$ & 11.98$\pm{0.15}^{\downarrow}$ & 66.11$\pm{0.83}$ & 74.96$\pm{0.61}^{\downarrow}$ & 59.51$\pm{1.08}$ & 62.74$\pm{0.98}$  \\
                           & Article & 11.27$\pm{0.12}^{\downarrow}$ & 8.61$\pm{0.12}^{\downarrow}$  & 15.01$\pm{0.2}^{\downarrow}$ & 20.68$\pm{0.33}^{\downarrow}$ & 18.19$\pm{0.16}^{\downarrow}$ & 18.86$\pm{0.23}^{\downarrow}$ \\
                           & Video-S & \textbf{3.99$\pm{0.14}$} & 3.7$\pm{0.15}$ & \textbf{5.29$\pm{0.21}$} & \textbf{8.97$\pm{0.34}$} & \textbf{5.31$\pm{0.18}$} & \textbf{6.2$\pm{0.23}$}  \\
                           & Video-L & 5.42$\pm{0.29}^{\downarrow}$ & 3.65$\pm{0.15}^{\downarrow}$ & 24.62$\pm{1.22}^{\downarrow}$ & 32.84$\pm{1.29}^{\downarrow}$ & 18.71$\pm{1.21}^{\downarrow}$ & 21.35$\pm{1.24}^{\downarrow}$  \\
\hline                           
\multirow{5}{*}{\textbf{CAT-ART}}   & APP-Ins & \textbf{38.36$\pm{0.58}$} & \textbf{27.96$\pm{0.31}$} & \textbf{24.86$\pm{0.34}$} & \textbf{35.46$\pm{0.39}$}  & \textbf{43.47$\pm{1.23}$} & \textbf{41.55$\pm{0.94}$} \\
                                    & APP-Use & \textbf{21.23$\pm{0.18}$} & \textbf{12.33$\pm{0.18}$} & \textbf{66.53$\pm{0.65}$} & \textbf{75.66$\pm{1.02}$}  & \textbf{59.98$\pm{0.86}$} & \textbf{63.27$\pm{1.02}$}  \\
                                    & Article & \textbf{16.82$\pm{0.21}$} & \textbf{12.4$\pm{0.13}$}  & \textbf{18.76$\pm{0.56}$} & \textbf{25.47$\pm{0.6}$}   & \textbf{25.97$\pm{0.61}$} & \textbf{25.79$\pm{0.58}$}  \\
                                    & Video-S & 3.93$\pm{0.08}$  & \textbf{3.93$\pm{0.06}$}           & 3.83$\pm{0.50}$ & 7.35$\pm{0.82}$ & 3.93$\pm{0.14}$ & 5.05$\pm{0.24}$ \\
                                    & Video-L & \textbf{6.08$\pm{0.09}$}  & \textbf{3.96$\pm{0.08}$}  & \textbf{27.18$\pm{0.39}$} & \textbf{35.01$\pm{0.67}$}  & \textbf{21.03$\pm{0.38}$} & \textbf{23.54$\pm{0.86}$}  \\
\bottomrule
\end{tabular}
\label{table:main_result}
}
\end{center}
\end{table*}

\subsection{Model Implementation and Complexity}
\textbf{Environment}. We implement our model using PyTorch~\cite{Pytorch} with python 3.6 and train the framework on Tesla P40 GPUs with a memory size of 22.38 GiB and a 1.53 GHz memory clock rate. 

\textbf{Proposed Method\footnote{\url{https://github.com/Chain123/CAT-ART}}}.
We use the BPRMF~\cite{rendle2012bpr} model with the user and item embedding size of $m=64$ for single domain recommendation and pretraining of domain-specific user embeddings. 
We reformulate the contrastive loss function given in~\cite{chen2020simple} for the contrastive autoencoder task in our model.  
For the CAT module, we build the autoencoder with Multi-Layer Perceptron~\cite{gardner1998artificial} where the sizes of the hidden layers are set to be $[5\times m, 3\times m, m]$ and $[m, 3\times m, 5\times m], m=64$ for the encoder and decoder modules, respectively. 
We adopt the PReLU activation function introduced by~\citeauthor{he2015delving}~\cite{he2015delving} between layers.  
We use a minibatch size of $N=4096$ in the training of the CAT module, and set $\tau = 0.1$ in the contrastive loss~\eqref{eq:contrastive_loss}. For each user, we mask out the domain-specific user embedding of one randomly selected domain out of the five domains in our collected dataset. 
In the loss function $\mathcal{L}_{cat}$, defined in~\eqref{eq:total_loss}, we set $\alpha_1 = \alpha_2 = 0.4$. 
The ART units in each domain have an identical structure, in which all adaptation modules are MLP layers with only one hidden layer which has the same size as its input, and the attention module is with only one head. 

\textbf{Model Complexity.} The time complexity of the model training is $\mathcal{O}(\max_{i \in [1, n]}I_i*d + |U|nd^2)$ where $n$ is the number of domains, $I_i$ is the number of interactions in domain $i$, $|U|$ is the number of users and $d$ denotes the embedding dimension of user.

\subsection{Experimental Results}
\label{sec:main_result}
Table~\ref{table:main_result} summarizes the performances of our proposed CAT-ART model and all the baseline methods on the collected multi-target CDR scenario with 5 domains. 
The proposed CAT-ART model outperforms all the other baselines in most of the domains. 
CAT-ART achieves the best performance in ``APP-Ins'', ``'APP-Use', ``Article'', and ``Video-L'' domains (\textbf{ RQ1}). 
CAT-ART also avoids negative transfer and outperforms most of the baselines in ``Video-S''. 
HeroGRAPH-L achieves a slightly better performance on the ``Video-S''. 
\blue{The reasons for this phenomenon are twofold. First, the user behaviors in the ``Video-S'' domain are richer and more diverse which makes it less likely to be affected by information from other domains, that is, it is hard to improve the recommendation performance in ``Video-S'' through cross-domain information, as is shown in Table~\ref{table:main_result}.}
Therefore, the recommendation performance in ``Video-S'' mainly depends on how we model user and item embeddings from the diverse user behavior data in the single domain. 
\blue{Secondly, we use the state-of-the-art graph-based recommendation model, i.e., Lightgcn~\cite{he2020lightgcn}, in the HeroGRAPH-L method, which is a deep Graph Convolutional Networks (GCN) that can generate better user and item embeddings in single domain compared to the MF model~\cite{wang2019neural}.}
In summary, HeroGRAPH-L can get a slightly better performance than our method in the ``Video-S'' domain due to the superiority of GCN over the MF model. 
However, HeroGRAPH-L still suffers from the negative transfer problem causing great performance reduction in other domains, such as ``APP-Use'', ``Video-L''. 
\blue{Furthermore, CAT-ART outperforms all the baselines in the other four domains.}

Compared with the SMF, we can see that CAT-ART effectively handles the negative transfer problem in all domains (\textbf{RQ2}). 
We attribute the success of the CAT-ART framework to the architecture design of the CAT and ART, which are specifically structured to avoid the negative transfer issue, while try to integrate as much useful information as possible from other domains to boost performance. 
Specifically, the CAT module generates high-quality global user embeddings for recommendations in all domains. And the ART module adoption of the attention mechanism to integrate domain-specific embeddings from other domains. We give a detailed analysis of each module in ablation studies in the following subsection. 
On the other hand, all the baseline cross-domain methods are severely affected by the negative transfer problem, causing significant performance degradation in many domains, e.g., ``APP-Use'', ``Articles'', and ``Video-L''. 
This is reasonable. First, as shown in Table~\ref{table:dataset_info}, the ``APP-Use'' and ``Video-L'' domains are much sparser than the other domains. 
Furthermore, user behavior in ``App-Ins'', ``APP-Use'' and ``Article'' are monotonous in nature compared with ``Video-S'' (therefore, the SMF achieves a much higher precision score in these domains). Obviously, when data from all domains are collected and trained together, domains with more sparsity tend to be overwhelmed by other domains causing biased cross-domain representation and negative transfer, e.g., in domains ``APP-Use'' and ``Video-L''. In addition, domains with simple and monotonous data prone to be affected by too much information in other domains, e.g., for domains ``App-Ins'', ``APP-Use'', and ``Articles''.  Apparently, according to the experimental results, none of the previous work can handle these situations(\textbf{ RQ2}). 

\begin{table*}[tbp]
\begin{center}
\caption{Results (in \%) of ablation studies. The $\downarrow$ represents negative transfer compared with the SMF model.} 
\resizebox{0.76\textwidth}{!}{
\begin{tabular}{ll|cccc|c}
\toprule
Domain                     & Metric       & SMF & +Autoencoder  & +Contrastive & +ART    &  -Attention \\  
\midrule
                           & Precision@10 & 33.82$\pm{0.70}$  & 37.64$\pm{1.17}$  & 37.95$\pm{0.45}$    & \textbf{38.36$\pm{0.58}$}  & 36.24$\pm{0.26}$  \\ 
                           & Recall@10    & 21.51$\pm{0.39}$  & 24.35$\pm{0.76}$  & 24.58$\pm{0.35}$    & \textbf{24.86$\pm{0.34}$}  & 23.35$\pm{0.23}$   \\
\multirow{-3}{*}{App-Ins}  & NDCG@10      & 32.56$\pm{0.43}$  & 41.34$\pm{3.75}$  & 42.56$\pm{2.02}$    & \textbf{43.47$\pm{1.23}$}  & 36.08$\pm{1.54}$  \\
\cline{2-7} 
                           & Precision@10 & 20.91$\pm{0.23}$  & 21.00$\pm{0.11}$  & 21.08$\pm{0.23}$    & \textbf{21.23$\pm{0.18}$}  & 21.01$\pm{0.07}$   \\ 
                           & Recall@10    & 65.50$\pm{0.89}$  & 65.77$\pm{0.33}$  & 66.01$\pm{0.88}$    & \textbf{66.53$\pm{0.65}$}  & 65.92$\pm{0.41}$  \\
\multirow{-3}{*}{APP-Use} & NDCG@10      & 57.39$\pm{1.46}$  & 59.09$\pm{0.37}$  & 58.61$\pm{0.40}$    & \textbf{59.98$\pm{0.86}$}   & 59.28$\pm{0.24}$   \\
\cline{2-7} 
                           & Precision@10 & 16.02$\pm{0.73}$  & 16.54$\pm{0.46}$  & 16.46$\pm{0.34}$    & \textbf{16.82$\pm{0.21}$}  & $15.88\pm{0.15}^{\downarrow}$    \\  
                           & Recall@10    & 16.64$\pm{0.43}$  & 17.48$\pm{1.21}$  & 17.19$\pm{1.13}$    & \textbf{18.76$\pm{0.56}$}  & 15.89$\pm{0.38}^{\downarrow}$    \\
\multirow{-3}{*}{Article}  & NDCG@10      & 21.59$\pm{1.30}$  & 23.98$\pm{2.28}$  & 23.54$\pm{2.75}$    & \textbf{25.97$\pm{0.61}$}  & 22.25$\pm{1.71}$ \\
\cline{2-7} 
                           & Precision@10 & 3.89$\pm{0.025}$  & 3.91$\pm{0.08}$   & \textbf{3.97$\pm{0.13}$} & 3.93$\pm{0.08}$  & 3.82$\pm{0.28}^{\downarrow}$ \\ 
                           & Recall@10    & 3.59$\pm{0.44}$   & 3.71$\pm{0.40}$   & 3.72$\pm{0.37}$          & \textbf{3.83}$\pm{0.50}$  & 3.46$\pm{0.25}^{\downarrow}$  \\
\multirow{-3}{*}{Video-S}  & NDCG@10      & 3.83$\pm{0.13}$   & 3.87$\pm{0.08}$   & 3.91$\pm{0.05}$          & \textbf{3.93}$\pm{0.14}$  & 3.73$\pm{0.18}^{\downarrow}$ \\
\cline{2-7} 
                           & Precision@10 & 5.98$\pm{0.20}$   & 6.04$\pm{0.01}$   & 6.07$\pm{0.04}$     & \textbf{6.08$\pm{0.09}$}   & 5.86$\pm{0.03}^{\downarrow}$  \\ 
                           & Recall@10    & 26.73$\pm{0.87}$  & 27.00$\pm{0.08}$  & 27.17$\pm{0.20}$    & \textbf{27.18$\pm{0.39}$}  & 26.27$\pm{0.09}^{\downarrow}$    \\
\multirow{-3}{*}{Video-L}  & NDCG@10      & 20.37$\pm{1.19}$  & 21.00$\pm{0.14}$  & 21.12$\pm{0.21}$    & \textbf{21.03$\pm{0.38}$}  & 20.26$\pm{0.15}^{\downarrow}$    \\
\bottomrule
\end{tabular}
\label{table:ablation_result}
}
\end{center}
\end{table*}

\subsection{Ablation Study and Analysis}
\red{We conduct ablation studies to show the effectiveness of each proposed module and to demonstrate how negative transfer is addressed through model design (\textbf{RQ3}). }
We incrementally accommodate different modules into the single-domain matrix factorization model (SMF), until we incorporate all the proposed sub-modules and features. Specifically, the following models are evaluated:
\begin{itemize} [leftmargin=*]
    \item \textbf{SMF}: The single-domain Matrix Factorization (MF) model.
    \item \textbf{+Autoencoder}: We add the original autoencoder to extract global representations for CDR.
    \item \textbf{+Contrastive}: We further add the contrastive loss for the training of the autoencoder, i.e., the CAT module.
    \item \textbf{+ART}: The ART module is further incorporated to integrate domain-specific user embedding.
\end{itemize}
Furthermore, to show the impact of negative transfer when domain-specific embeddings are directly transferred without attention, we further remove the attention layer within the ART module. 
Specifically, the following model is evaluated:
\begin{itemize}[leftmargin=*]
    \item \textbf{-Attention}: We remove the attention from the ART and only use MLP layers to integrate domain-specific features.
\end{itemize}

Table~\ref{table:ablation_result} summarizes the results of ablation studies. 
The values of metrics Precision@10, Recall@10, and NDCG@10 are given.
We can see, each time we add a new sub-module or feature incrementally on top of the previous model, we can observe an improvement on the overall recommendation performance, which illustrates the effectiveness of autoencoder, contrastive self-supervised learning, and the ART modules. 
Specifically, 
with the original autoencoder framework, we already get rid of the negative transfer in all domains of our dataset. This can be attribute to the natural advantage of the autoencoder framework where the encoder is trained to extract the most important information to reconstruct the input, thus, is able to reduce the effect of noisy sample. 
\red{Furthermore, by adding contrastive training (\emph{+Contrastive}), we improve the performance in most of the domains via a more general user representation. However, the recommendation is not always improved, especially in the ``Article'' domain. The reasons are twofold. 1) In our work, the goal of contrastive learning is to make the user embedding more robust to noise and less dependent on domain-specific information. 2) The ``Article'' domain is less related to the rest of the domains, that's why all the baselines methods encounter the negative transfer problem in this domain, shown in Table~\ref{table:main_result}. Therefore, a more domain-independent global embedding given by the CAT module has less domain-specific information from the ``Article'' domain, resulting in slightly worse recommendation performance compared to the user embedding given by the autoencoder.}
Finally, domain-specific embeddings
are incorporated through the \emph{ART} module to further boost recommendation performance, while avoiding negative transfer through attention. 
\red{Note that, 
due to the special data characteristics in the ``Video-S'' domain, we can only get a relative small improvement. Thus, all variants have close performance. However, it is clearly shown in Table~\ref{table:ablation_result} that we get the best performance when incorporating all the modules, i.e., \emph{+ART}, in the rest of the domains.}

\begin{figure}[tbp]
  \centering
  \includegraphics[width=2.3in]{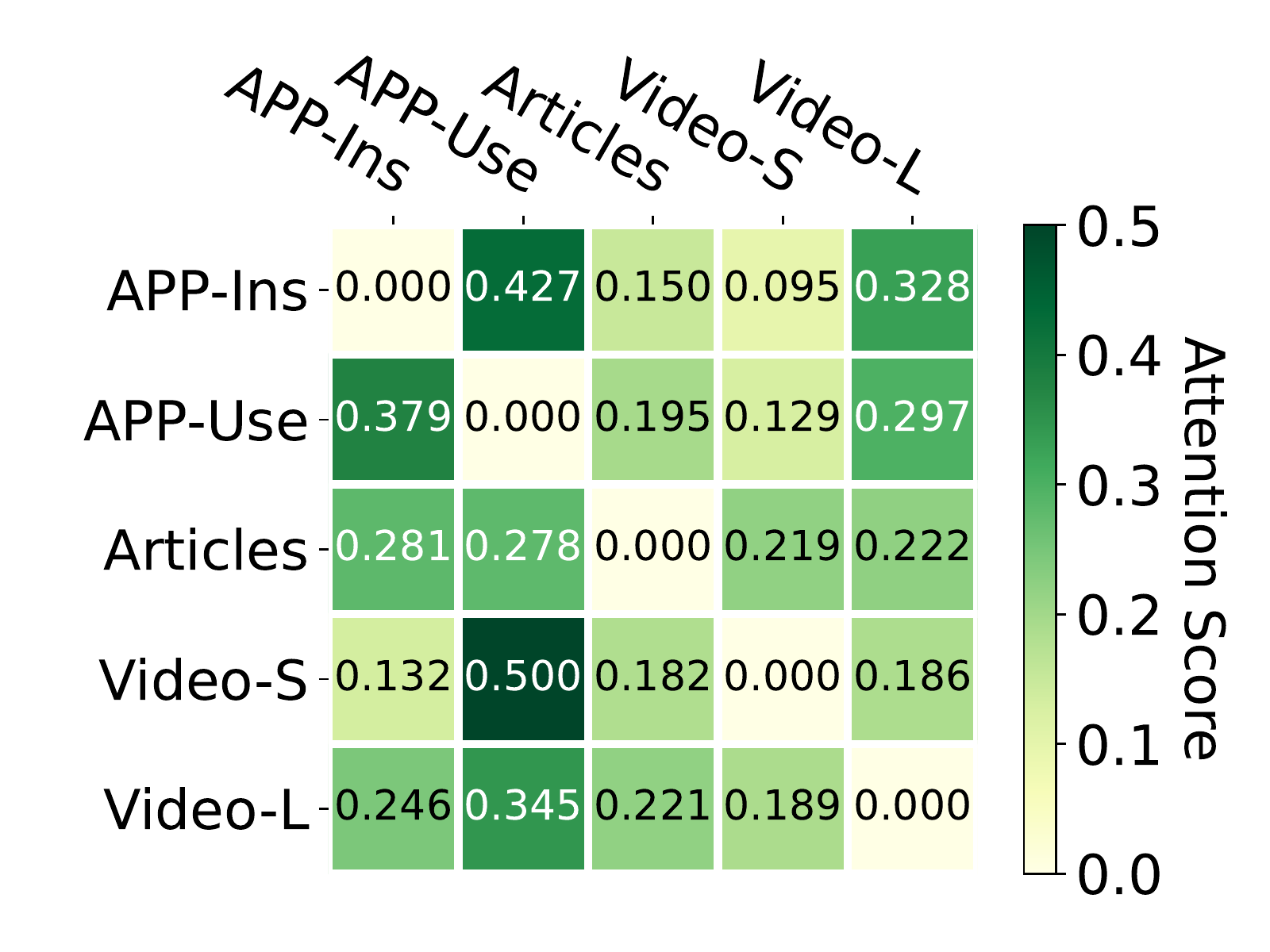}
  \caption{Averaged attention scores on the test set. 
  }
  \Description{A figure shows the overall structure of the CAT-ART model.}
  \label{fig:attention_score}
\end{figure}

Furthermore, without attention (\emph{-Attention}), the performance is greatly reduced in all domains and the negative transfer problem also prevails, which demonstrate the effectiveness of the attention module in avoiding the negative transfer problem.   

Figure~\ref{fig:attention_score} shows the averaged attention scores on the test set, given by the ART module in each individual domain. Each row represents the attention weights assigned to the source domains by the corresponding target domain, therefore the sum of weights in a row equals 1. 
We can see, the weights between related domains are high, e.g., domains ``APP-Ins'' and ``APP-Use'', while weights between unrelated domains are low, e.g., ``APP-Ins'' and ``Video-L''. 
Furthermore, the asymmetry of the weight matrix shows the inequality in the use of shared domain embedding between two domains. 
For example, ``APP-Use'' assigns a weight of $0.13$ to the ``Video-S'' domain, while the ``Video-S'' gives an attention score of $0.5$ to the ``APP-Use''. 
This is reasonable, it's helpful for the recommendation of videos if we know what Application a user is more likely to use, but the other way around is much harder. 
These phenomena further demonstrate the need for a mechanism to select the most important and helpful information among the features provided by multiple domains.

\section{Conclusion}
\label{sec:conclude}

In this paper, we focus on the MTCDR problem and propose the CAT-ART model.  
We build a CAT module to extract robust unbiased global user representation in a self-supervised manner via contrastive learning and an autoencoder framework based on pretrained domain-specific user embeddings.
Then, the ART module is built in each domain, which transfers domain-specific user embeddings from other domains with the attention mechanism. 
Combining these two modules, CAT-ART boosts recommendation in all participating domains and avoiding negative transfer at the same time. 
We believe CAT-ART has made a valuable contribution in exploring the MTCDR and the negative transfer issue, approaching the objectives of One for All and All for One. 

\bibliographystyle{ACM-Reference-Format}

\end{document}